\documentclass[12pt]{iopart}

\usepackage[dvipdfm]{graphicx}
\usepackage{color}

\begin{document}

\title{
Nonlinear MHD simulation of core plasma collapse events in stellarators
}

\author{Yasuhiro Suzuki$^{1,2}$, Shimpei Futatani$^3$, and Joachim Geiger$^4$}

\address{$^1$ National Institute for Fusion Science, National Institutes of Natural Sciences, 509-5292 Toki, Japan}
\address{$^2$ The Graduate University for Advanced Studies, SOKENDAI, 509-5292 Toki, Japan}
\address{$^3$ Polytechnic University of Catalonia, 08034 Barcelona, Spain}
\address{$^5$ Max-Planck-Institut f\''{u}r Plasmaphysik, 17491 Greifswald, Germany}

\ead{suzuki.yasuhiro@nifs.ac.jp}

\pacno{52.55.Hc, 52.55.-s, 52.30.Cv, 52.30.-q, 52.25.Fi}

\begin{abstract}

Three-dimensional nonlinear MHD simulations study the core collapse events observed in a stellarator experiment. In the low magnetic shear configuration like the Wendelstein 7-X, the rotational transform profile is very sensitive to the toroidal current density. The 3D equilibrium with localized toroidal current density is studied. If the toroidal current density follows locally in the middle of the minor radius, the rotational transform is also changed locally. Sometimes, the magnetic topology changes due to appearing the magnetic island. A full three-dimensional nonlinear MHD code studies the nonlinear behaviors of the MHD instability. It was found that the following sequence. At first, the high-$n$ ballooning-type mode structure appears in the plasma core, and then the mode linearly grows. The high-$n$ ballooning modes nonlinearly couple and saturate. The mode structure changes to the low-$n$ mode. The magnetic field structure becomes strongly stochastic into the plasma core due to the nonlinear coupling in that phase. Finally, the plasma pressure diffuses along the stochastic field lines, and then the core plasma pressure drops. This is a crucial result to interpret the core collapse event by strong nonlinear coupling.  

\end{abstract}

\maketitle


\section{Introduction}

The stellarator is a magnetic configuration to confine the fusion plasma. In the stellarator, the rotational transform is produced by the external coils only, and then the magnetic field configuration is intrinsically three-dimensional (3D). In such a sense, the nonlinearity of the magnetic configuration is very strong due to the nonlinear coupling of the modes.

In the stellarator, the disruption like the tokamak~\cite{Hender2007} usually does not happen, but the core collapse event, which is not terminating the plasma, sometimes happens in the experiment. For example, in the Large Helical Device (LHD) experiment, the core density collapse~\cite{Sakamoto2007}, so-called CDC, ocuurs in the high-central beta plasma with the steep pressure gradient~\cite{Ohyabu2006}. If the CDC happens, the core electron density, not the core electron temperature, drops with a fast time scale. Some physical interpretations were proposed, for example, the internal reconnection event and so on. However, since the precursor of the CDC is localized in the weak field side, where the magnetic curvature is bad, it was considered that the ballooning mode drives the precursor, particularly the resistive ballooning mode~\cite{Ohdachi2010,Ohdachi2017}. Recently, in another stellarator experiment of the Wendelstein 7-X (W7-X), the core collapse was found in the current drive experiment by the electron cyclotron current drive (ECCD)~\cite{Wolf2017,Zanini2020}. In the core collapse event, the core electron temperature sharply drops like the tokamak sawtooth. Some theoretical predictions were proposed~\cite{Zocco2019}, but the mode was not experimentally validated. 

The 3D nonlinear MHD simulation code is a fundamental to understand the nonlinear behavior of the stellarator plasmas. Many 3D nonlinear simulations were examined in the stellarator plasmas, and then the importance of the nonlinearity has been clarified~\cite{Miura2004,Miura2006,Miura2007,Miura2010,Sato2012,Miura2014,Ichiguchi2014,Ichiguchi2015,Miura2017,Sato2017,Futatani2019,Schlutt2012,Roberds2016}. The 3D nonlinear MHD simulation was examined to understand the CDC~\cite{Mizuguchi2009}. In the nonlinear simulation, it was found that the ballooning mode appeared in the middle of the minor radius, the mode stochastized the magnetic field structure by the nonlinear coupling. After that, the plasma pressure diffused along the stochastic field line, and the core plasma pressure dropped due to the pressure diffusion. The interesting point is that the ballooning mode in the middle or edge plasma drives the core collapse due to the nonlinear coupling~\cite{Mizuguchi2009}.     

In this study, the core collapse event observed in the W7-X is studied by a full 3D nonlinear MHD simulation code, MIPS (MHD Infrastructure for Plasma Simulation)~\cite{Todo2010}. In the next section, the 3D equilibrium, including the localized toroidal current density, is studied. Since the localized toroidal current density strongly affects the rotational transform profile with the low magnetic shear, the rotational transform and magnetic field structure changes are studied. And then, the nonlinear MHD simulation initialized by the 3D equilibrium field is examined. Finally, this study is summarized.

\newpage

\section{Impacts of localized toroidal current density on 3D equilibrium}

In previous studies, it was found that the localized toroidal current density, $\textbf{j}_{\phi}$, significantly affects the magnetic topology~\cite{Suzuki2016}. Here, the impacts of the localized toroidal current density are studied systematically with the assumed toroidal current density profile.

\begin{figure}[htbp]
 \begin{center}
  \includegraphics{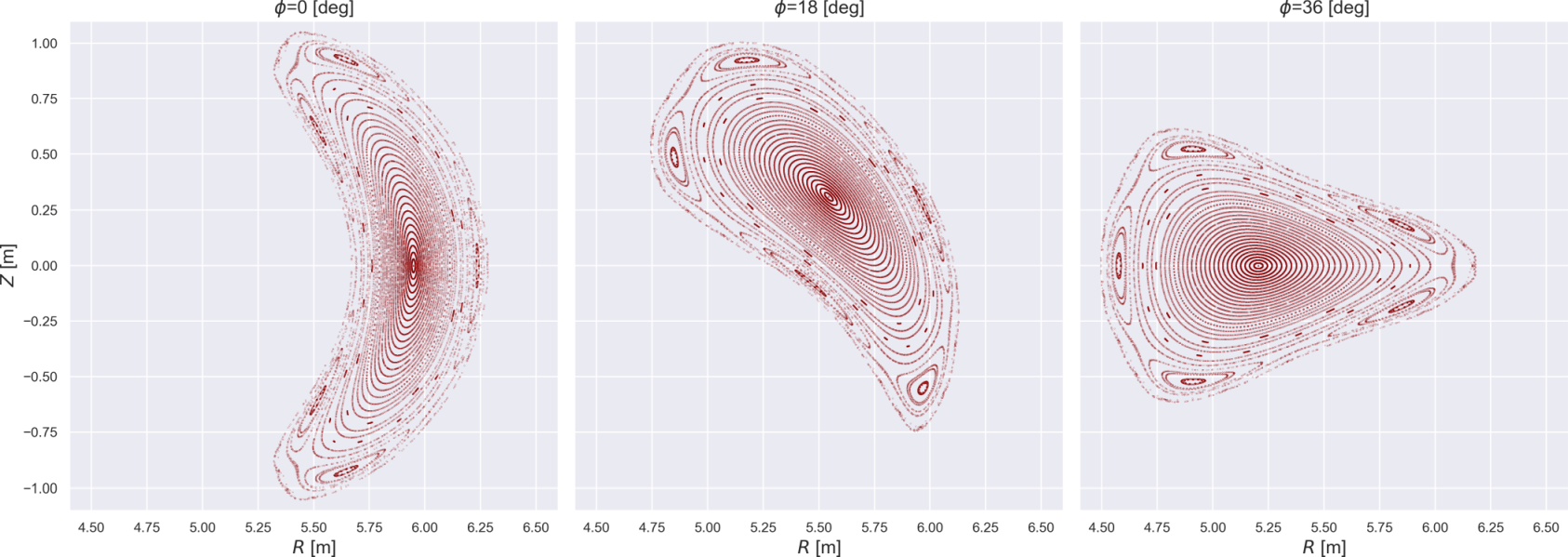}
 \end{center}
 \caption{Poincar\'e plot of the vacuum magnetic field for the standard configuration is shown at three poloidal cross sections, $\phi$ = 0, 18, and 32 deg.}
 \label{fig:vacuum}
\end{figure}

In \fref{fig:vacuum}, Poincar\'e plots of the standard configuration are shown for the vacuum magnetic field at three different poloidal cross sections, $\phi$ = 0, 18, and 36 degrees. Clear flux surfaces appear in the inside of an island chain on an $\iota$=1 rational surface, and the last closed flux surface (LCFS) is prescribed by the separatrix of 5/5 islands of the $\iota$=1.

The 3D equilibrium is analyzed by HINT code, which is a 3D equilibrium calculation code without an assumption of nested flux surfaces~\cite{Suzuki2006,Suzuki2017,Suzuki2020}. Thus, the topological change of the magnetic field structure, in particular, the magnetic island or stochastization of the magnetic field, can be analyzed. In the HINT modeling, the magnetic field is initialized by the vacuum magnetic field. And, the plasma pressure and toroidal current density profiles are given as inputs, and fixed in the equilibrium calculation. The pressure profile is approximated by a parabolic function as,
\begin{equation}\label{eq:p-profile}
    p = p_0 (1-s),
\end{equation}
where $p_0$ is the pressure on the axis, and $s$ is the normalized toroidal flux. The toroidal current density profile is approximated by a Gaussian function as,
\begin{equation}\label{eq:j-profile}
    j_{\phi} \propto a \exp \left ( \frac{- \left( s - b \right )^2 }{c^2} \right ),
\end{equation}
where parameters, $a=1$, $b=0.2$, and $c=0.05$ are used. That means the toroidal current density profile of \eref{eq:j-profile} is localized around $s$ = 0.2. The plasma beta is defined by,
\begin{equation}
   \beta _0 = \frac{2 \mu_0 p_0}{{B_0}^2},
\end{equation}
where $B_0$ is the magnetic field on the axis at $\phi$ = 0. In this study, the beta value, $\beta_0$ is fixed to 1\%. The toroidal current density is prescribed by the net toroidal current,
\begin{equation}
    I_{\phi} = \int j_{\phi} dS.
\end{equation}
In this study, the net toroidal current is scanned by $I_{\phi}$ = 0, 10, 15, and 20 kA with the fixed beta value. The field periodicity is assumed, that is, the equilibrium calculation is conducted for only one field period, from 0 to 72 deg.

\begin{figure}[htbp]
 \begin{center}
  \includegraphics{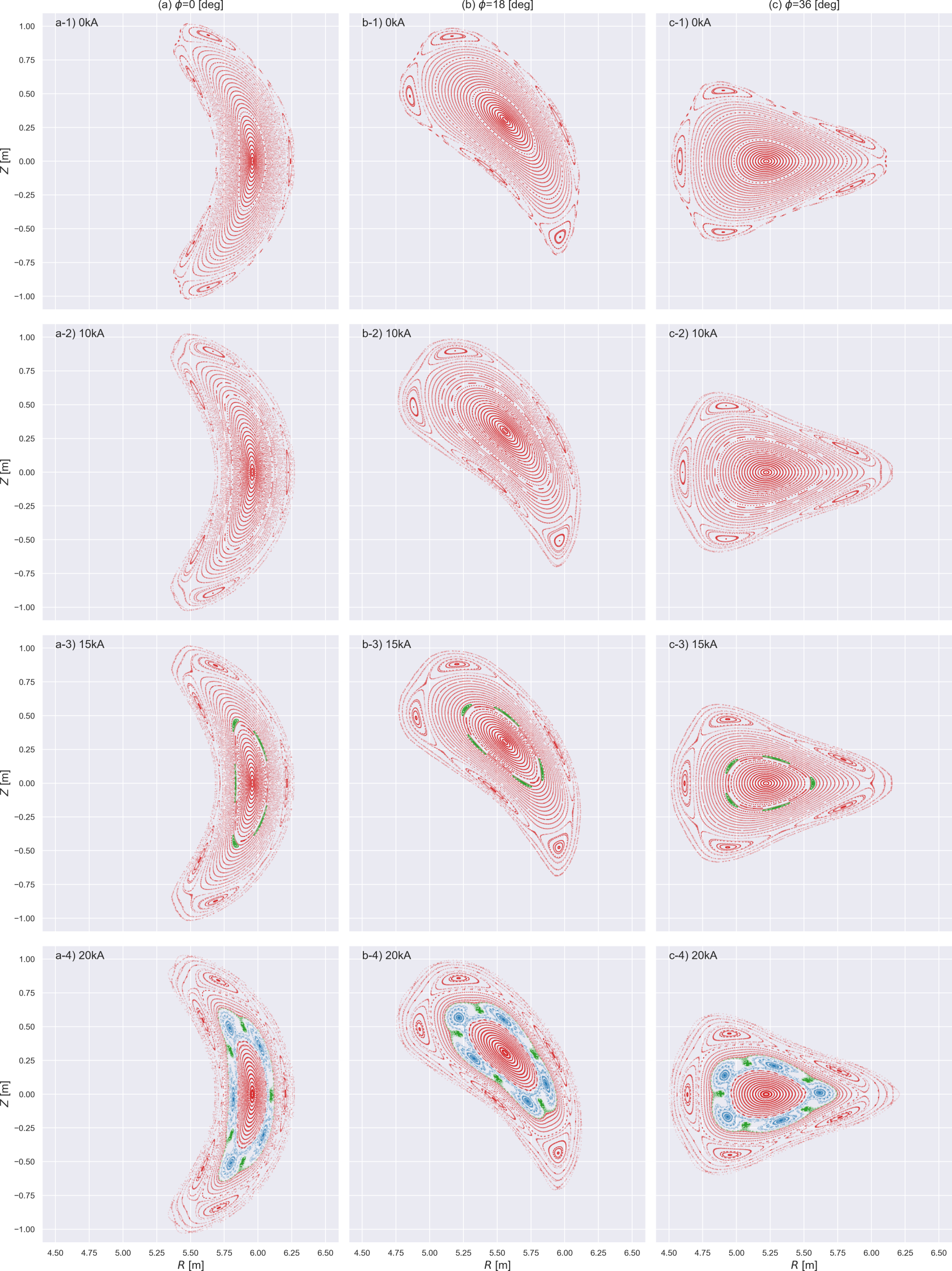}
 \end{center}
 \caption{Poincar\'e plots of the 3D equilibrium analyses are shown for the sequence of the net toroidal current scan. Poloidal cross sections of Poincar\'e plots are corresponding to \fref{fig:vacuum}. Rows in the figure are corresponding to cases of $I_{\phi}$ = 0, 10, 15, and 20 kA.}
 \label{fig:equilibrium}
\end{figure}

\begin{figure}[htbp]
 \begin{center}
  \includegraphics{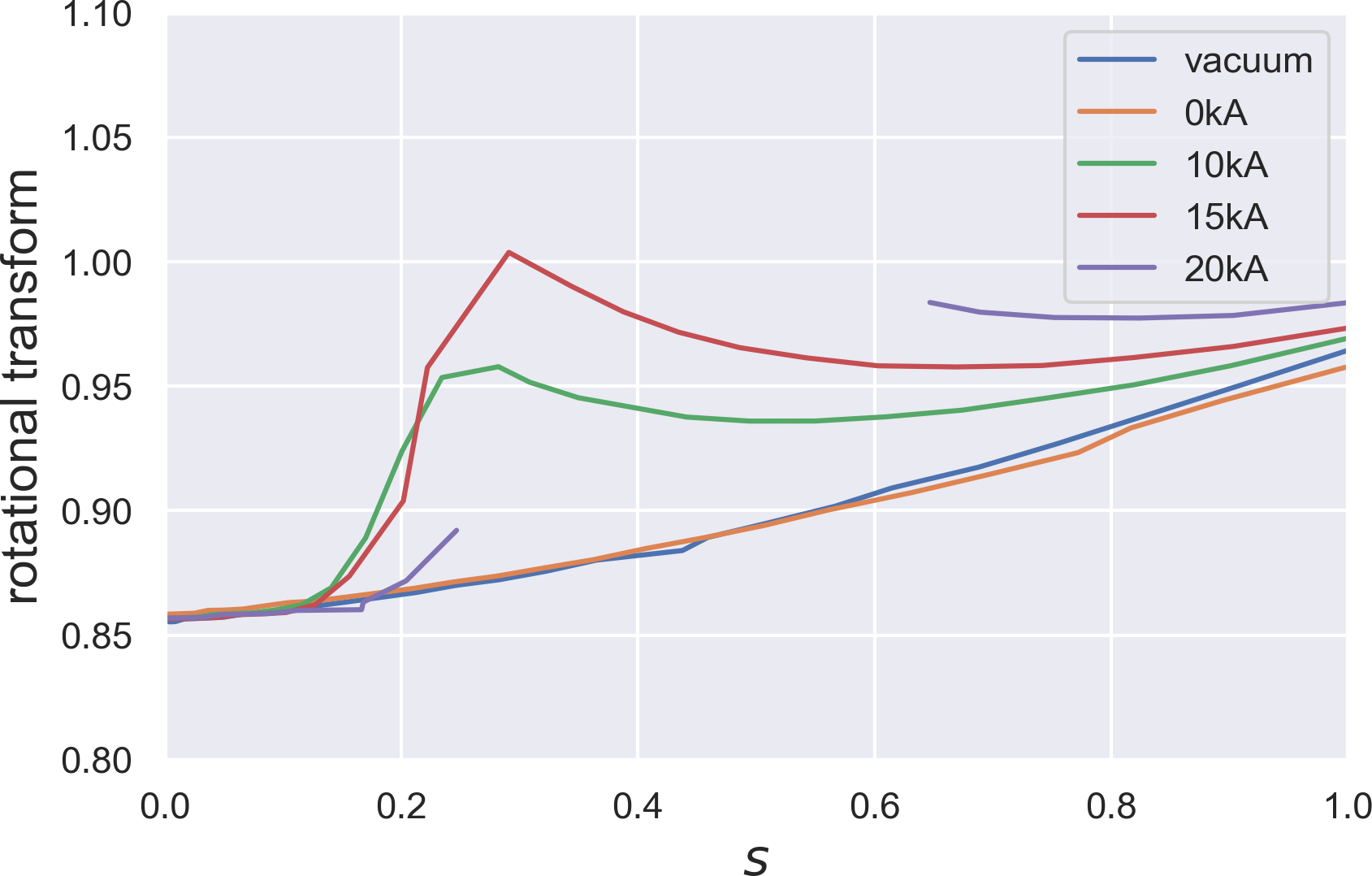}
 \end{center}
 \caption{Rotational transform profiles of the 3D equilibrium analyses for the sequence of the net toroidal current scan.}
 \label{fig:iota}
\end{figure}

In \fref{fig:equilibrium}, Poincar\'e plots of the 3D equilibrium analysis are shown for the sequence of the net toroidal current scan. Poloidal cross sections of Poincar\'e plots are corresponding to \fref{fig:vacuum}. Rows in the figure are corresponding to cases of $I_{\phi}$ = 0, 10, 15, and 20 kA. For $I_{\phi}$ = 0 and 10 kA, no significant differences do not appear. Clear flux surfaces are kept in the inside of $\iota$ = 1, and 5/5 islands also do not appear. However, for $I_{\phi}$ = 15 kA, a new island chain of 5/5 islands (green) appears in the plasma core. This is caused by the localized toroidal current density, which locally increases the rotational transform around $s \sim$ 0.2. Furthermore, for $I_{\phi}$ = 20 kA, another island chain of 5/5 islands (blue) appears. This means that the rotational transform crosses the $\iota$ = 1 twice. To understand these significant changes of the magnetic topology, the sequence of the rotational transform for $I_{\phi}$ = 0, 10, 15, and 20 kA is shown in \fref{fig:iota}. For reference, the vacuum rotational transform is also shown. Rotational transform profiles for the vacuum and $I_{\phi}$ = 0 kA are almost identical, because the plasma beta is low in this study. However, for $I_{\phi}$ = 10 kA, the rotational transform increases locally around $r_\mathrm{eff} \sim$ 0.25. For $I_{\phi}$ = 15 and 20 kA, the rotational transform increased larger than the $\iota$ = 1. For $I_{\phi}$ = 15 kA, the rotational transform achieves to the $\iota$ = 1 with the thin region, and then the small island chain appears. On the other hand, for $I_{\phi}$ = 20 kA, the rotational transform becomes $\iota >$ 1 in the thick region. Thus, the new island chain (blue) with the out of the original phase (green) appears. The interesting point is that how these local changes of the rotational transform affects the nonlinear evolution of the MHD stability. That is discussed in the next section.

\newpage

\section{Impacts of localized iota changes on nonlinear MHD instability}

\subsection{Model equations and numerical setup}

In this study, a full 3D nonlinear MHD simulation code, MIPS (MHD Infrastructure for Plasma Simulation)~\cite{Todo2010}, is utilized. The MIPS code computes the nonlinear dissipative MHD equations starting from the initial equilibrium; in the MHD-approximation, the plasma is modeled as a charge-neutral electro-magnetic conducting fluid. In the incompressible description, the full-set of equations is given by
\begin{eqnarray}
\frac{\partial \rho}{\partial t} + \nabla \cdot \left ( \rho \mathbf{v} \right ) = \nabla \cdot \left ( D_{\perp} \rho \right ), \\
\rho \left (\frac{\partial \mathbf{v}}{\partial t} + \mathbf{v} \cdot \nabla \mathbf{v} \right ) = -\nabla p + \mathbf{j} \times \mathbf{B} + \frac{4}{3} \left[ \nu \rho \nabla \cdot \mathbf{v} \right ] - \nabla \times \left ( \nu \rho \mathbf{\omega} \right ), \\
\frac{\partial p}{\partial t} + \nabla \cdot \left ( p \mathbf{v} \right ) + \left ( \Gamma - 1 \right ) p \nabla \cdot \mathbf{v} = \nabla \cdot \left ( \chi_{\perp} \nabla p \right ) + \nabla \cdot \left [ \mathbf{B} \left ( \frac{\chi_{\parallel}}{B^2} \left ( \mathbf{B} \cdot \nabla \right ) p \right ) \right ], \\
\frac{\partial \mathbf{B}}{\partial t} = -\nabla \times \mathbf{E}, \\
\mathbf{E} + \mathbf{v} \times \mathbf{B} = \eta \left ( \mathbf{j} - \mathbf{j}_\mathrm{eq} \right ),
\end{eqnarray}
where the vorticity $\mathbf{\omega} = \nabla \times \mathbf{v}$, $p$ is the plasma pressure. These equations are solved on the cylindrical coordinates ($R, Z, \phi$). The resistivity $\eta$, the viscosity $\nu$ and the perpendicular and parallel heat conductivities $\chi_{\perp}$ and $\chi_{\parallel}$ affect as dissipation parameters in the equations. The system is normalized by $ v_A R_0$, where $v_A$ is the Alfv\`en speed and $R_0$ is a major radius. The equilibrium current density $\mathbf{j}_\mathrm{eq}$ corresponds to the Pfirsch-Schl\"uter (P-S) current and net toroidal current density. In this simulation, the plasma beta, $\beta_0$, is low, and the impact of the P-S current may be small.

In the MIPS code, all physical quantities are discretized on the rectangular grid of the cylindrical coordinate. The size of the simulation box along $R$ and $Z$ directions is the square box of 2.2 m by 2.2 m. The number of grid points is used to (128,128,320). The spatial difference is approximated by the fourth-order central finite difference scheme, and the fourth-order Runge-Kutta scheme is used for the time integration. However, since the numerical oscillation sometimes happens in the approximation of the convection terms, the Kawamura-Kuwabara scheme (third-order upwind scheme) is used for the convection terms in order to avoid numerical oscillation~\cite{Kawamura1984}. The equilibrium magnetic field, current density, and pressure distribution are initialized by the HINT code discussed in the previous section. The mass density distribution is calculated by a adiabatic equation, $\rho = C p^{-\Gamma}$, where $C$ is a constant.

The resistivity is approximated by the Spitzer resistivity, that is, the resistivity is proportional to $T_e^{-\frac{3}{2}}$. Thus, the profile of the resistivity is assumed to
\begin{equation}
    \eta \propto \eta_0 T_e^{-\frac{3}{2}},
\end{equation}
where $\eta_0 = {10}^{-8}$. The electron temperature is calculated from the relation, $p = 2 k n_e T_e$. Here, the $k$ is the Boltzmann constant and the $T_e = T_i$, $Z_\mathrm{eff}$ = 1 are assumed. The perpendicular and parallel heat conductivities are assumed to be the constant, $\chi_{\perp} = {10}^{-6}$ and $\chi_{\parallel}$ = 0. In the presented simulations, the pressure diffusion parallel to the magnetic field has been neglected by setting $\chi_{\parallel}$ = 0. Also, the viscosity is the constant, $\nu = {10}^{-6}$. 

\subsection{nonlinear simulation results}

The localized net toroidal current significantly affects the rotational transform, in other words, the magnetic field. An open question is that how the localized change of the magnetic field affects the nonlinear MHD behavior.

\begin{figure}[htbp]
 \begin{center}
  \includegraphics{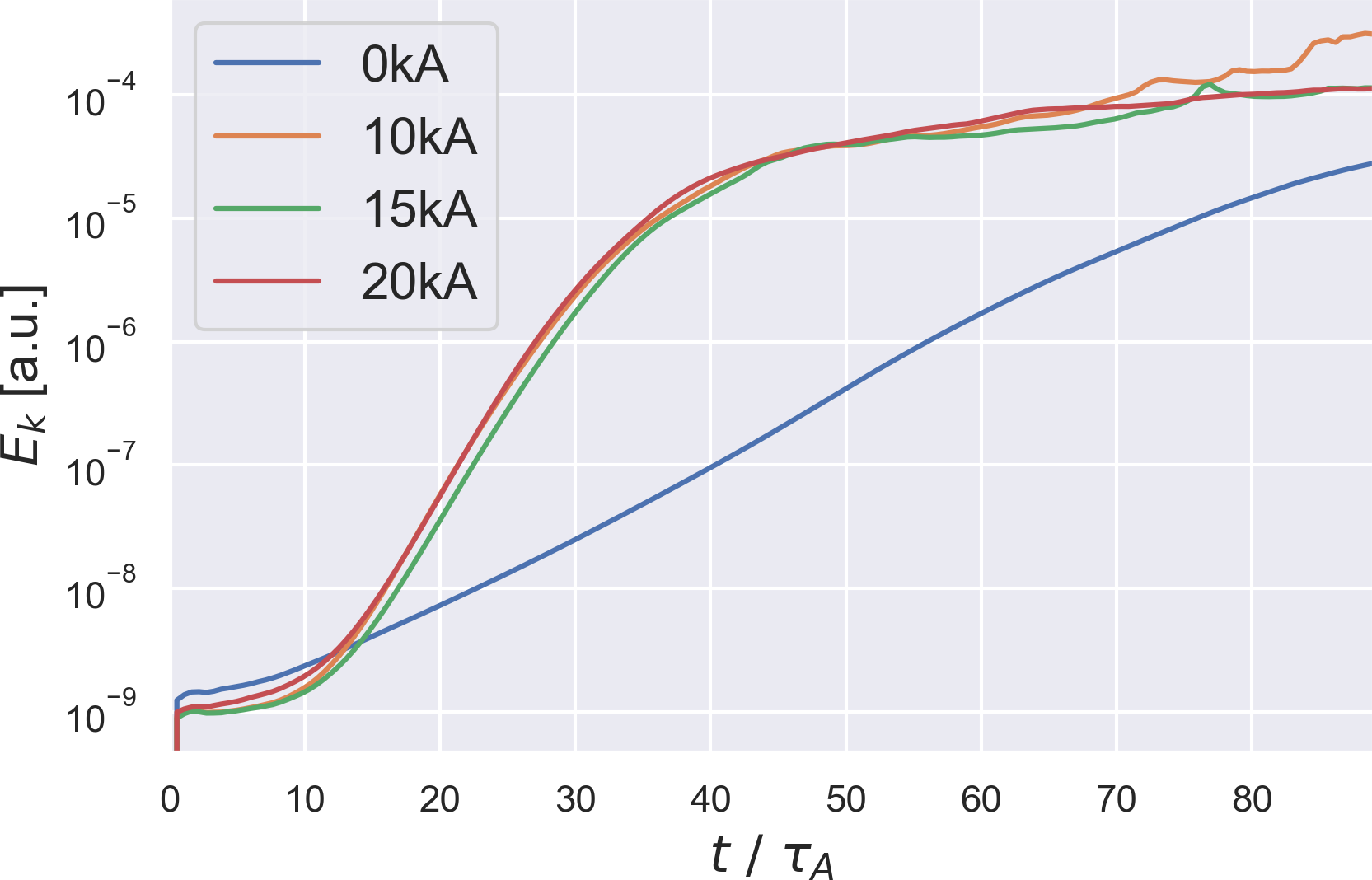}
 \end{center}
 \caption{The time evolution of nonlinear MHD simulations for the sequence of the net toroidal current scan.}
 \label{fig:Ek}
\end{figure}

\Fref{fig:Ek} shows the time evolution of the kinetic energy for the sequence of the net toroidal current from 0 to 20 kA. For cases with the net toroidal current from 10 to 20 kA, the energy rapidly grows linearly, and then nonlinearly saturates. For cases of 10 and 20 kA, the linear growth are almost same, and for another case of 15 kA, the linear growth is slightly small. However, the linear growth of 0 kA is significantly small comparing with cases including the net toroidal current. In \tref{tab:gamma}, the growth rate, $\gamma / \omega_\mathrm{A}$ calculated from the linear growth of $E_k$ is shown for all cases. The $\gamma / \omega_\mathrm{A}$ of 10 kA is largest, but $\gamma / \omega_\mathrm{A}$ of 15 kA is small although the rotational transform achieved to the unity. The $\gamma / \omega_\mathrm{A}$ of 20 kA becomes large comparing with the case of 15 kA.

\begin{figure}
 \begin{center}
  \begin{tabular}{cc}
   \includegraphics{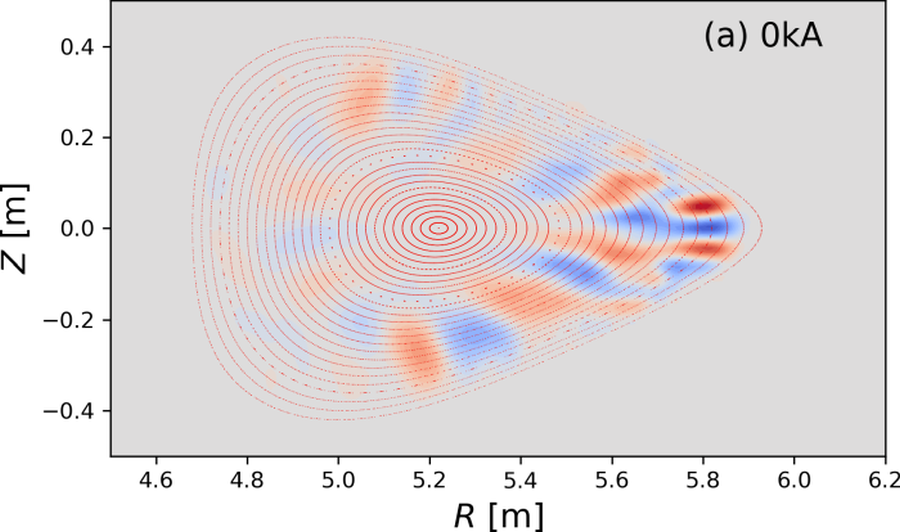} &
   \includegraphics{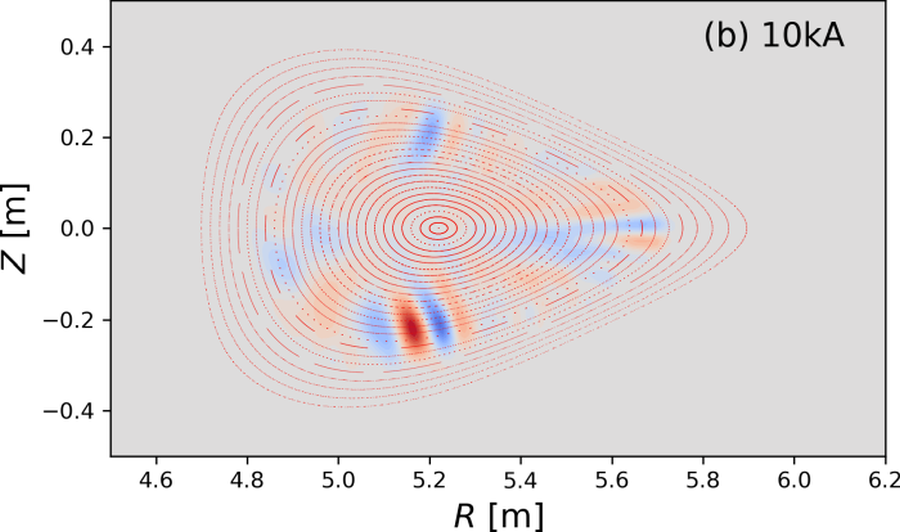} \\
   \includegraphics{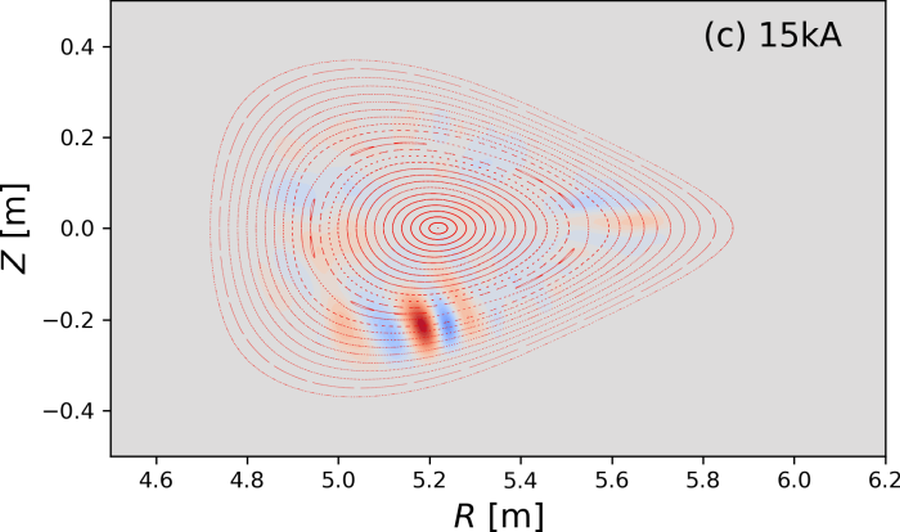} &
   \includegraphics{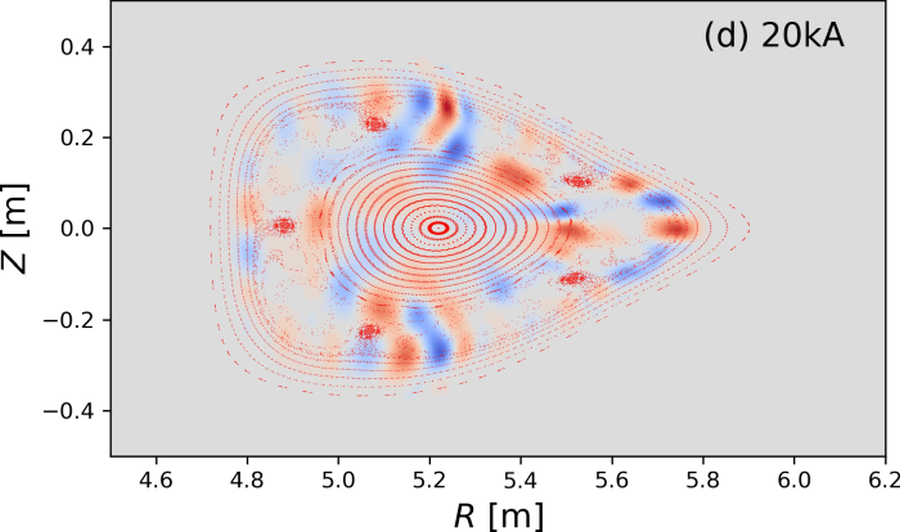}
  \end{tabular}
 \end{center}
 \caption{Mode structures of the linearly growing phase for different net toroidal currents.}
 \label{fig:linear}
\end{figure}

\Fref{fig:linear} shows the mode structure of the linear growth for all cases. For reference, Poiancar\'e plots of the 3D equilibrium field corresponding to each cases are also shown. A largest mode of 0 kA appears in the edge region, but, for other cases of 10 and 15 kA, largest mode appears in more inside of the plasma core. A hypothesis appearing mode in the plasma core is the magnetic shear. In \fref{fig:iota}, the localized toroidal current increases the rotational transform at $s > 0.2$, and the magnetic shear becomes very weak in there. In such a case, the mode may be appeared in the region of the weak magnetic shear. For the last case of 20 kA, the modes are more complicated. The mode appears strongly in the edge region, because the magnetic shear becomes very weak. On the other hand, the mode also appears strongly in the inside of 5/5 island chains. This is probably caused by the distortion of the pressure profile due to large magnetic islands. 

\begin{figure}
 \begin{center}
  \begin{tabular}{ccc}
  (a) magnetic field &
  (b) perturbed pressure &
  (c) pressure \\
   \includegraphics{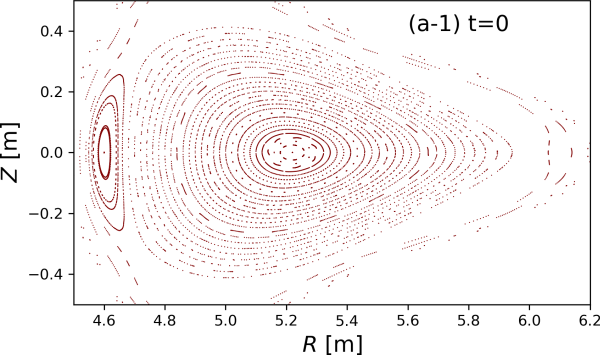} &
   \includegraphics{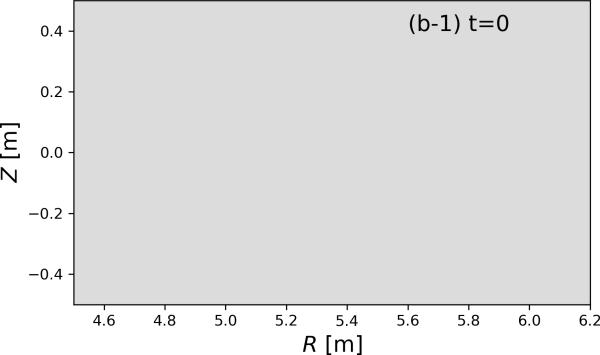} &
   \includegraphics{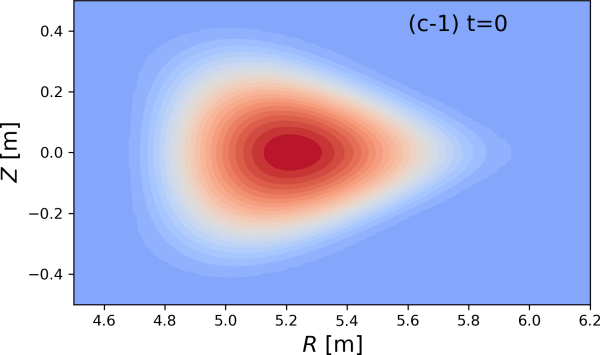} \\
   \includegraphics{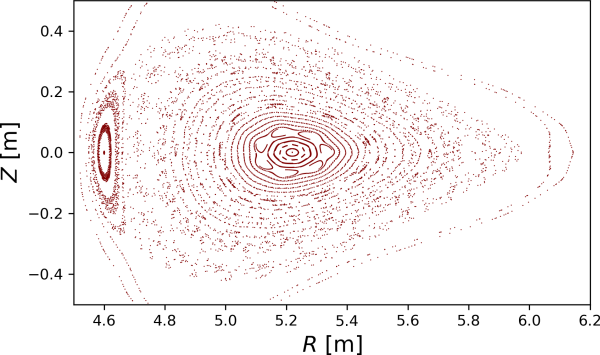} &
   \includegraphics{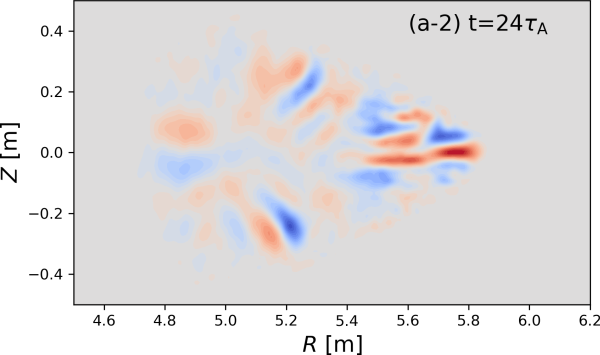} &
   \includegraphics{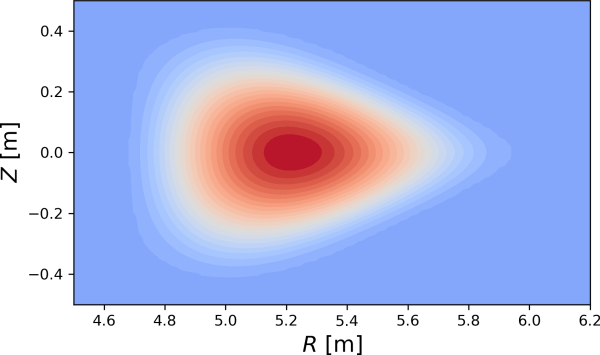} \\
   \includegraphics{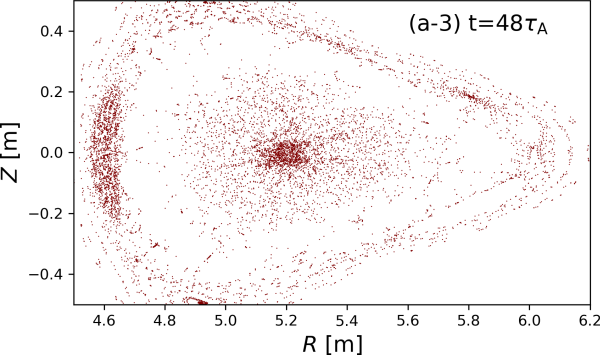} &
   \includegraphics{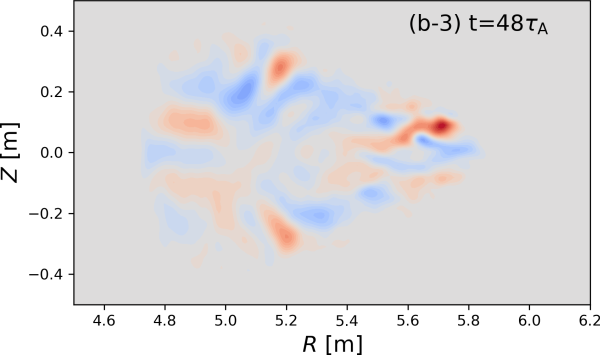} &
   \includegraphics{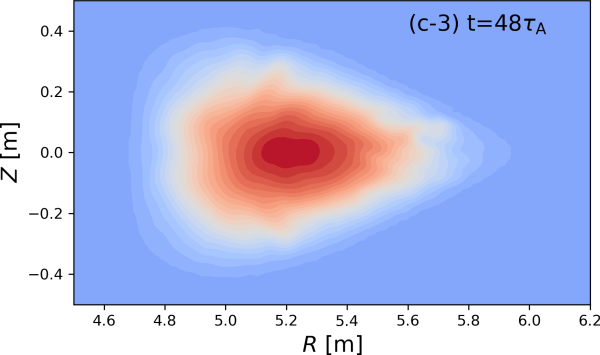} \\
   \includegraphics{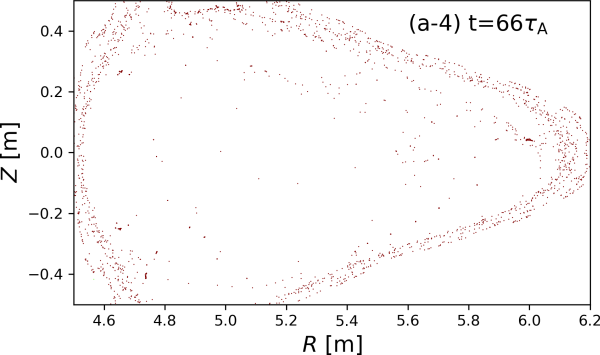} &
   \includegraphics{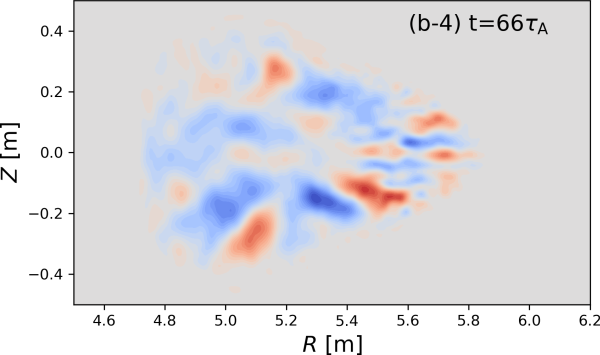} &
   \includegraphics{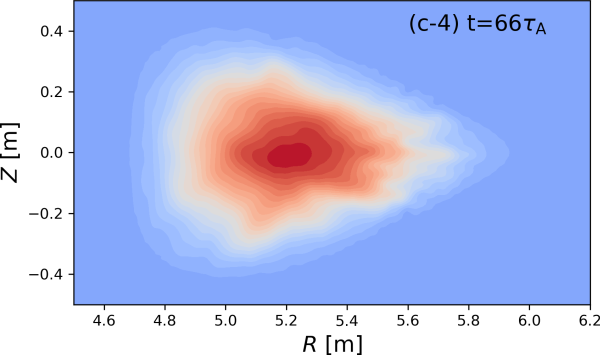} \\
   \includegraphics{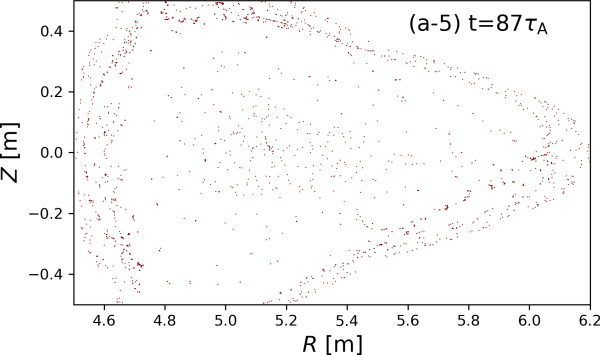} &
   \includegraphics{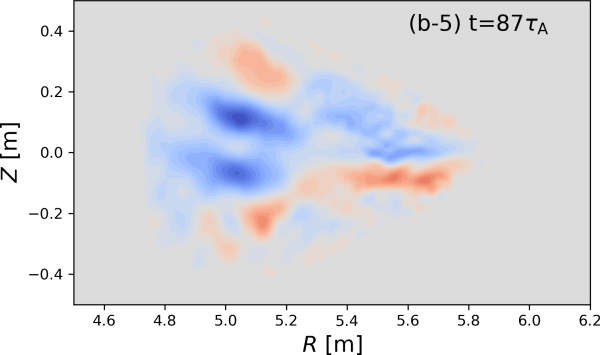} &
   \includegraphics{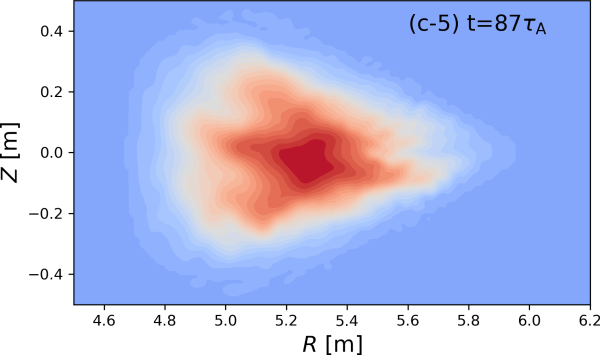}
  \end{tabular}
 \end{center}
 \caption{The time evolution of (a) the magnetic field, (b) perturbed pressure, $\tilde{p}$, and (c) pressure, $p$ for a case of 10 kA. Rows in the figure are corresponding to cases of $t$ = 0, 24 $\tau_\mathrm{A}$, 48 $\tau_\mathrm{A}$, 66 $\tau_\mathrm{A}$, and 87 $\tau_\mathrm{A}$. }
 \label{fig:nonlinear}
\end{figure}

It is found that the linear growth rate is largest for the case of 10 kA. Thus, the nonlinear behavior is studied for the case of 10 kA. \Fref{fig:nonlinear} shows the nonlinear evolution of (a) the magnetic field, (b) perturbed pressure, $\tilde{p}$, and (c) pressure, $p$. In figures, five time slices at (1) $t = 0 \tau_\mathrm{A}$, (2) $t = 24 \tau_\mathrm{A}$, (3) $t = 48 \tau_\mathrm{A}$, (4) $t = 66 \tau_\mathrm{A}$, and (5) $t = 87 \tau_\mathrm{A}$. At $t = 24 \tau_\mathrm{A}$, the linear mode grows sufficiently, and the mode penetrates into the core. The magnetic field in the core still keeps nested flux surfaces, but, in the edge, the stochastization begins slightly. According to the small stochastization, the pressure distribution keeps the almost equilibrium profile. After the nonlinear saturation at $t = 48 \tau_\mathrm{A}$, the magnetic field becomes strongly into the core. The distortion of the pressure distribution begins due to the nonlinearly saturated mode. At $t = 66 \tau_\mathrm{A}$, the magnetic filed is completely broken. An important point is that the mode structure changes from high-$n$ to low-$n$ modes due to the nonlinear coupling. The pressure deformation becomes strong and the diffusion of the pressure along the stochastic field begins. For the final phase at $t = 87 \tau_\mathrm{A}$, the magnetic field is still stochastic. The mode structure changes to the low-$n$ modes further, and the pressure deforms and diffuses. These results suggest that the mode in the linear phase appears as the high-$n$ modes in the edge but, in the nonlinear phase, the modes nonlinearly saturate and couple. And then, the mode structure changes to the low-$n$ mode in the core.   

Finally, a remaining question that how these nonlinear behaviors reflect the core collapse is considered. \Fref{fig:p-evolve} shows the nonlinear evolution of the pressure profile on the triangular cross section ($\phi$ = 32 deg) is shown as the function of $t\ /\ \tau_\mathrm{A}$. In the linear phase, the deformation of the pressure begins from the edge according to the linear mode structure. However, in the nonlinear phase, the pressure profile deforms into the core, and the diffusion of the pressure begins. Therefore, the peak pressure reduced due to the large diffusion of the pressure along the stochastic field line. These behaviors is very similar to the CDC observed in the LHD experiment. And also, this nonlinear simulation suggests that the core collapse may not be driven by the core mode directly. The strong nonlinear coupling may cause the low-$n$ mode structure to collapse the plasma core.

\newpage

\section{Summary and Discussion}\label{sc:summary}

This study deals with the 3D nonlinear MHD modeling of the core collapse event in a stellarator, W7-X. At first, the 3D equilibrium, which includes the localized toroidal current density modeling the ECCD, is studied by the HINT code. The rotational transform profile is very sensitive to the localized toroidal current density. Depending on the net toroidal current, the rotational transform locally increases and achieves the unity for $I_\mathrm{net} >$ 15 kA. In such cases, the 5/5 magnetic island chain appears at a new rational surface of $\iota$ = 1 in the plasma core. In particular, for the case of 20 kA, double island chains of 5/5 appear, and phases of double island chains are opposite each other. Next, the 3D nonlinear MHD is studied by the MIPS code. For the net toroidal current free, the linear growth is very small. However, for cases with the net toroidal current, the mode linearly grows, the ballooning type mode structure appears in the plasma core. The weak magnetic shear causes that due to the increased rotational transform. After the linear growth, the modes nonlinearly are coupled and then saturated. In the nonlinear saturation phase, the mode structure changes from the high-$n$ modes to the low-$n$ modes. At the same time, the magnetic field becomes stochastic. That causes the diffusion of the plasma pressure along the stochastic field line, and the core pressure is collapsed. That is the hypothesis of the core collapse driven by the nonlinear coupling of the high-$n$ ballooning mode. According to this hypothesis, the core collapse event may not be directly driven by the low-$n$ mode on the $\iota$ = 1 surface, for example, $m/n$ = 1/1 mode.

\begin{figure}[htbp]
 \begin{center}
  \includegraphics{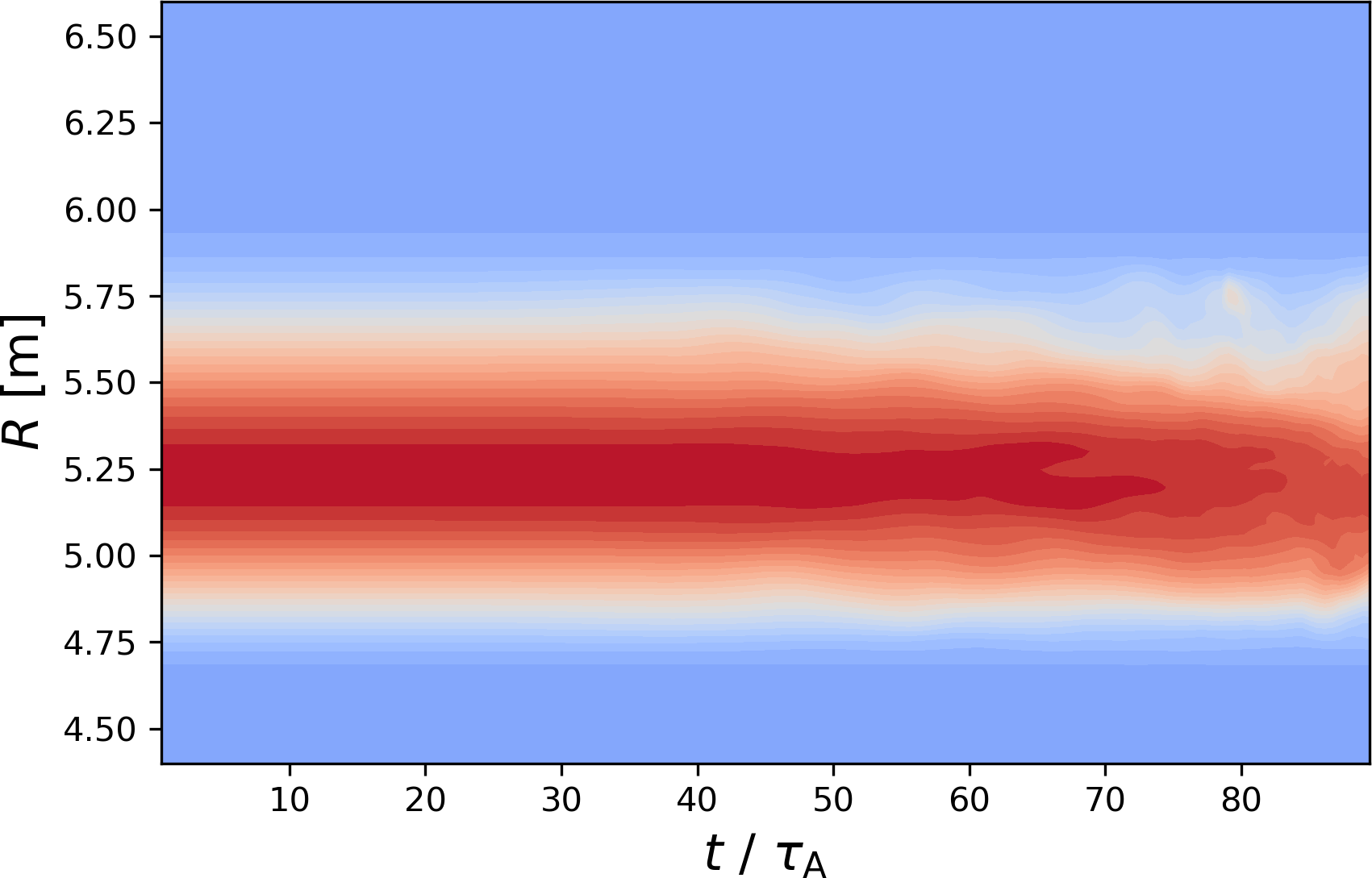}
 \end{center}
 \caption{The time evolution of the plasma pressure for the case of 10 kA. The profile is shown on $Z$ = \textit{const}. plane at the triangular plane, $\phi$ = 32 deg.}
 \label{fig:p-evolve}
\end{figure}

A remaining discussion is the fast parallel transport along the stochastic field line. In this study, the parallel heat conductivity, $\chi_{\parallel}$ is set to zero. Since the MIPS code uses the explicit time integration scheme, it is inefficient to involve the large ratio of $\chi_{\parallel} / \chi_{\perp}$ in the simulation. Therefore, in this study (see \fref{fig:p-evolve}), although the magnetic field becomes stochastic at the beginning of the nonlinear saturation phase, the plasma pressure does not flatten. To flatten the plasma pressure on the stochastic field immediately, the ratio of $\chi_{\parallel} / \chi_{\perp}$ should be $10^{6} \sim 10^{8}$. However, to involve $\chi_{\parallel} / \chi_{\perp}$ of $10^{6} \sim 10^{8}$, the time step of the integration for the MHD equations must be sufficiently small or the parallel diffusion term must be integrated implicitly. An implicit solver of the parallel diffusion term is now developing. The result, including the large ratio of $\chi_{\parallel} / \chi_{\perp}$, is a future subject, and it will be presented in a different paper. 

\newpage

\ack
The author (Y.S.) greatly acknowledges instructive discussions with Prof. K.~Ichiguchi (NIFS). This work is performed on ``Plasma Simulator'' (NEC SX-Aurora TSUBASA) of NIFS with the support and under the auspices of the NIFS Collaboration Research (NIFS18KNST130,  NIFS20KNST171) and was performed on the JFRS-1 supercomputer system at Computational Simulation Centre of International Fusion Energy Research Centre (IFERC-CSC) in Rokkasho Fusion Institute of QST. This work is supported by JSPS (the Japan Society for the Promotion of Science) Grant-in-aid for Scientific Research (B) 18H01202, and the NINS (National Institute of Natural Sciences) program of Strategic International Research Interaction Acceleration Initiative (Grant Number UFEX404). Also, this work was partially supported by ``PLADyS'', JSPS Core-to-Core Program, A. Advanced Research Networks.

\noappendix




\newpage

\begin{table}[htpb]
 \begin{center}
  \begin{tabular}{c|cccc}
   $I_\mathrm{ECCD}$ & 0 kA & 10 kA & 15 kA & 20 kA \\ \hline
   $\gamma / \omega_\mathrm{A}$ &  0.45 & 0.64 & 0.53 & 0.60 
  \end{tabular}
 \end{center}
 \caption{The linear growth rate for cases of different $I_\mathrm{ECCD}$.}
 \label{tab:gamma}
\end{table}

\end{document}